\documentclass[twocolumn,           % Format : preprint, twocolumn
               showpacs,            % Pacs : showpacs, noshowpacs
               preprintnumbers,     % Preprint: preprintnumbers,
               aps,                 % Society: ...
               prd,                 % Journal Style : pra, prb, prc, prd, pre,
               a4paper,             % Size : a4paper, ...
               superscriptaddress,  % Affiliation (Title) : groupedaddress,
               nofootinbib,         % Footnote: footinbib, nofootinbib
               tightenlines,        % Remove additional spaces in a line
               floats,floatfix      % Floating pictures and tables
               ]{revtex4}

\usepackage{graphicx}
\usepackage{subfigure}
\usepackage{latexsym}
\usepackage{amsmath,amssymb}        % amssymb includes amsfonts
\usepackage[draft=false]{hyperref}
\usepackage{threeparttable}
\usepackage{multirow}

\input colordvi

\begin{document}

%%--- DRAFTCOPY --------------------------------
%% Prints a large "DRAFT" diagonally across each page
%% Does not show up in TeXview
%% \typeout{Prints "DRAFT" on each page; does not show in TeXView}
%% \special{!userdict begin /bop-hook{gsave 200 30 translate
%% 65 rotate /Times-Roman findfont 216 scalefont setfont
%% 0 0 moveto 0.90 setgray (DRAFT) show grestore}def end}
%%------------------------------------------------

%======================================%
%<<<<<<<<<<<< TITLE PAGE >>>>>>>>>>>>>>%
%======================================%

\title{Constraining the dark fluid}
\author{Martin Kunz}
\affiliation{Astronomy Centre, University of Sussex, Brighton BN1 9QH,
  United Kingdom}
\author{Andrew R.~Liddle}
\affiliation{Astronomy Centre, University of Sussex, Brighton BN1 9QH,
  United Kingdom}
\author{David Parkinson}
\affiliation{Astronomy Centre, University of Sussex, Brighton BN1 9QH,
  United Kingdom}
\author{Changjun Gao}
\affiliation{The National Astronomical Observatories, Chinese Academy
  of Sciences, Beijing, 100012, China}
\date{\today}
\pacs{95.36.x,98.80.-k}

%======================================%
%<<<<<<<<<<<<< ABSTRACT >>>>>>>>>>>>>>>%
%======================================%

\begin{abstract}
Cosmological observations are normally fit under the assumption that
the dark sector can be decomposed into dark matter and dark energy
components. However, as long as the probes remain purely
gravitational, there is no unique decomposition and observations can
only constrain a single dark fluid; this is known as the dark
degeneracy. We use observations to directly constrain this dark fluid
in a model-independent way, demonstrating in particular that the data
cannot be fit by a dark fluid with a single constant equation of
state. Parameterizing the dark fluid equation of state by a variety of
polynomials in the scale factor $a$, we use current kinematical data
to constrain the parameters. While the simplest interpretation of the
dark fluid remains that it is comprised of separate dark matter and
cosmological constant contributions, our results cover other model
types including unified dark energy/matter scenarios.
\end{abstract}

\maketitle

%======================================%
%<<<<<<<<<<<<<< ARTICLE >>>>>>>>>>>>>>>%
%======================================%

\section{Introduction}

The standard cosmological model appeals to two separate dark
components --- dark matter and dark energy --- and the usual
application of observational constraints places limits on each of
these. However, at present these two components have only been
detected through their gravitational influence, and these measurements
do not provide enough information to permit a unique decomposition
into these components. Rather, it is a model assumption that the two
components are separate. This point was first made in
Ref.~\cite{hueis} and specific consequences explored in
Refs.~\cite{earlydeg1,earlydeg2}. Ref.~\cite{mydeg} extended it to
coupled models and named it the dark degeneracy.

This degeneracy is extremely general. One can write
\begin{equation}
G_{\mu\nu} - \kappa T_{\mu\nu}^{{\rm visible}} = \kappa T_{\mu \nu}^{{\rm dark}} \,
\end{equation}
where $G_{\mu\nu}$ is the Einstein tensor, $T_{\mu\nu}$ the
energy--momentum tensor, and $\kappa$ the gravitational coupling. Any
gravitational probe of the dark sector amounts to evaluating the
left-hand side of this equation, and interpretting the non-zero result
as evidence for the dark sector. Having obtained the dark sector
energy--momentum tensor this way, such observations can offer no
guidance on how that tensor might be split amongst different dark
components.\footnote{We work entirely in Einsteinian gravity. A somewhat
  related issue arises in modified gravity models, where the
  `non-Einsteinian' gravitational terms could be shifted to the
  right-hand side of the generalized Einstein equation and potentially
  reinterpretted as matter terms, see e.g.\ Ref.~\cite{demod}.}

Although the degeneracy holds even for structure formation probes
(both linear and non-linear) of the dark sector, we will focus here
only on kinematical probes, i.e.\ those referring to the homogeneous
and isotropic background cosmology. A complete dark sector description
is then given by the total dark sector energy density and its equation
of state $w_{{\rm dark}}$, the latter determining the evolution of the
former. This enables a simple analysis. To include structure formation
data, while retaining statistical homogeneity, one would also need to
consider at least the dark sector sound speed and perhaps also
anisotropic stress \cite{hueis,mydeg}, and one might expect that
structure formation observations would end up mostly constraining
their form rather than imposing further upon $w_{\rm dark}$.

As a simple example \cite{hueis}, in the standard cosmological model
the redshift evolution of the total dark sector equation of state
\begin{equation}
w_{{\rm dark}} \equiv \frac{\sum \rho_i w_i}{\sum \rho_i} \,,
\end{equation}
(where `$i$' runs over the dark components) is given by
\begin{eqnarray}
\label{e:lcdm}
w_{\rm dark}^{\rm SCM}(z) & = & - \frac{1-\Omega_{{\rm
 m},0}}{1-\Omega_{{\rm 
 m},0}+\left(\Omega_{{\rm m},0}-\Omega_{{\rm b},0}\right) (1+z)^3} \,,\\
 & \simeq & - \frac{1}{1+0.31 (1+z)^3} \,. \label{e:scm}
\end{eqnarray}
Here $\Omega_{\rm m}$ and $\Omega_{\rm b}$ are the total matter
density parameter and the baryon density parameter, and the subscript
indicates present value. The second line follows from inserting the
values $\Omega_{{\rm m},0} \simeq 0.274$ and $\Omega_{{\rm b},0} \simeq
0.046$ obtained from current data compilations \cite{Komatsu5yrWMAP}.

Inclusion of a single dark energy component with this equation of
state evolution would give the same observational predictions as the
standard cosmology. Indeed, once one allows the dark energy equation
of state to evolve arbitrarily, one cannot say anything from
observations about the dark matter density $\Omega_{\rm dm}$, as its
effects can always be reinterpretted as due to the dark
energy. Analyses which appear to measure $\Omega_{\rm dm}$ accurately
only manage to do so because the dark energy parameterization adopted
is not general enough to be able to mimic the form of
Eq.~(\ref{e:lcdm}).

This degeneracy is perfect. All we can say in a model-independent way
is that the present total dark sector density is about 0.95, and that
its equation of state is constrained to evolve in a particular way
from an early-time value at or near zero to arrive at its present
value of $w_{\rm dark}^{\rm obs}(z=0) \simeq -0.8$. Our aim in this
paper is to more precisely quantify these constraints, by considering
general dark sector models that do not include a pure cold dark matter
component.

\section{Models and Data}

Having set up this non-standard framework for analyzing the dark
sector, our analysis procedure is standard and straightforward.

\subsection{Models}

In order to impose constraints on the dark fluid, we need to employ a
parameterization of its equation of state. Henceforth we drop the
subscript `dark', using $w$ throughout as the total dark sector
equation of state. As we wish to reach high redshift, we parameterize
$w$ as a function of scale factor $a$, which has the bounded domain
$0<a\leq 1$. We simply expand this as a power series in $a$ at the
present, which we find to be sufficient. At linear order this is the
well-known CPL parameterization $w=w_0+(1-a)w_1$ \cite{CPL},
normally applied to the dark energy alone but here referring to the
combined dark sector. To general order this expression was given,
again for the dark energy alone, in Ref.~\cite{bck}, which in actual
calculation considered the linear and quadratic versions. As we are
demanding that our parameterization describe the entire dark sector, we
will explore up to cubic order. Our models are hence
\begin{equation}
w(a) = \sum_{n=0}^N w_n (1-a)^n
\end{equation}
for different choices of $N$.

As we will see, the data strongly demand that $w(a)$ is close to zero
as $a \rightarrow 0$, i.e.\ the dark sector is required to behave as
dark matter at early times (note that this is true even without
inclusion of any structure formation data). Accordingly we will also
consider the same expansions supplemented with the additional
constraint $w(a=0) = 0$ which fixes their highest-order term as a
function of the others, and with the combined constraints $w(a=0)=0$
and $\left.dw/da\right|_{a=0} = 0$. We will call these the
`constrained' and `doubly-constrained' expansions respectively. The
former can be used from linear order upwards, and the latter from
quadratic upwards.

In all cases we choose priors on the expansion coefficients which are
wide enough that the outcome is determined entirely by the data and
not the prior.

The standard rulers measured by the cosmic microwave background (CMB)
and baryon acoustic oscillations (BAO) are all fixed by the sound
horizon at decoupling (which we fix to be $z_{\rm dec}=1089$), which
is dependent on the baryon density.  The standard rulers are all
distance ratios independent of $H_0$, and the supernovae make no
measurement of the Hubble parameter $H_0$ since it is degenerate with
the unknown absolute normalization over which we marginalize. As our
data compilation does not constrain $H_0$, it is not able to give an
accurate constraint on the total dark sector energy density. However
the physical baryon density $\Omega_{\rm b} h^2$ is accurately
measured to the usual value, and so if supplemented with a direct
determination of $H_0$, we would find the expected total dark sector
energy density $\Omega_{{\rm dark}} \simeq 0.95$. In our analysis we
allow $\Omega_{\rm b} h^2$ and $H_0$ to vary, but constrain them using
other datasets, and marginalize over them, in effect marginalizing
over the total dark sector energy density.

In light of the above, when we quote model parameter counts they are
of the dark sector equation of state, and do not include the total
dark sector energy density. For the general dark sector models, we
quote the number of parameters specifying $w(a)$. For $\Lambda$CDM,
the equivalent parameter is the relative amount of dark matter and
dark energy at present (equivalently, the coefficient of the
\mbox{$(1+z)^3$} term in Eq.~(\ref{e:scm})).

\subsection{Data}

We use a fairly typical combination of kinematical data to constrain
our models. Standard candle data comes from supernova type Ia
luminosity distances, for which we use the cut Union supernova sample
\cite{Kowalski:2008ez} (with systematic errors included), and standard
ruler data comes from the angular positions of the CMB
\cite{Wang:2007mza} and BAO peaks \cite{Percival:2009xn}.  Note that
Ref. \cite{Wang:2007mza} give constraints on the scaled distance to
recombination $R$ and the angular scale of the sound horizon
$l_a$. These are defined to be
\begin{equation}
R \equiv \sqrt{\Omega_{\rm m} H_0^2}\, r(z_{\rm CMB}) \,, \quad l_a \equiv
\frac{\pi r(z_{\rm CMB})}{r_s(z_{\rm CMB})} \,. 
\end{equation}
Since $R$ is scaled by the physical matter density, and so makes
assumptions about the separability of the dark matter and dark energy,
we ignore it in this work. We use only the constraints on $l_a$, as
well as those on $\Omega_{\rm b} h^2$ and the correlations between the
two.  This still assumes that the sound speed of the dark component at
high redshift is small, in order to avoid an early ISW effect
shifting the peak position. We also include the SHOES (Supernovae and
$H_0$ for the Equation of State) measurement of the Hubble parameter
today, $H_0=74.2 \pm 3.6 {\rm ~kms}^{-1} {\rm Mpc}^{-1}$
\cite{Riess:2009pu}.

\section{Results}

\begin{table}
\caption{Parameters (of the dark sector equation of state) and
  best-fit chi-squared for our various models. The constrained models
  force $w(a=0)=0$, and the doubly-constrained ones additionally
  $\left. dw/da\right|_{a=0}=0$.}
\label{tab:modchi}
\begin{tabular}{lcc}
Model & ~~Dark sector~~ & $\chi^2_{{\rm min}}$ \\
 & parameters & \\%Total & SN  & CMB   & BAO\\
\hline
$\Lambda$CDM & 1 & 311.9 \\
Constant $w$ & 1 & 391.3 \\
Linear (CPL) & 2 & 312.1 \\
Constrained linear & 1 & 320.5 \\
Quadratic & 3 & 309.8 \\
Constrained quadratic & 2 & 311.9 \\
Doubly-constrained quadratic & 1 &  313.5 \\
Cubic & 4 & 309.6 \\
Constrained cubic & 3 & 311.1 \\
Doubly-constrained cubic &  2 & 311.5 \\
\hline
\end{tabular}
\end{table}

Table~\ref{tab:modchi} shows the number of adjustable parameters and
best-fit chi-squared for most models, including $\Lambda$CDM for
comparison. The total number of data points is 313 (308 SN-Ia, 2 BAO,
2 CMB and 1 from the SHOES project), but correlations between the data
points make it difficult to state the number of independent data
points, and so the number of degrees of freedom. We can say that the
number of degrees of freedom is, at most, 311 minus the number of dark
sector parameters, meaning that the data is an acceptable fit to all
the models, except the constant $w$ model, if the correlations are
small.

\begin{figure}[t]
\includegraphics[width=0.9\linewidth]{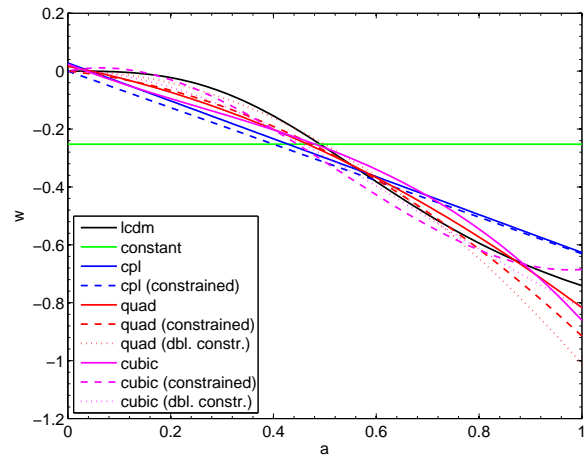}
\caption{The best-fit $w(a)$ for our various models. Note that the
  approach to $w=0$ at $a=0$ is determined entirely by the data in the
  unconstrained cases, while being enforced in the constrained models.}
\label{fig:wofa}
\end{figure}

The corresponding $w(a)$ curves of the best-fitting version of each
model are shown in Fig.~\ref{fig:wofa}. The immediate conclusions from
these are as follows:
\begin{enumerate}
\item $\Lambda$CDM, as expected, does a good job of fitting the data,
  bettered only by other models with more dark sector parameters. For
  our data the best-fit $\Omega_{\rm m}$ is $0.26 \pm 0.02$. This
  agrees well with the result of Komatsu et al.~\cite{Komatsu5yrWMAP},
  with somewhat larger uncertainty as less data is being used.
\item Even the best-fit version of the constant $w$ model is
  a very poor fit. The dark sector has not had a constant
  equation of state throughout its evolution.
\item The full four-parameter cubic does not significantly improve the
  fit over the three-parameter quadratic, indicating that
  three-parameter models saturate the constraining power of the data.
\item All the models have $w(a=0)$ at or very close to zero, enforced
  entirely by the data. This indicates that the full phenomenology can
  be captured using the constrained versions of the expansion,
  reducing the variable parameter set by at least one.
\item The doubly-constrained quadratic (i.e.\ simply $w(a) = w_0 a^2$)
  gives a tolerable one-parameter fit to the data, though not quite as good
  as the $\Lambda$CDM fit. By contrast, the single-parameter
  constrained linear fit is a poor fit to the data when compared to the
  $\Lambda$CDM fit which has the same number of parameters (though the general
  CPL model can acceptably fit it by overshooting to $w>0$ at early
  times).
\item In terms of giving good fits to the data for economical numbers
  of parameters, apart from $\Lambda$CDM the two-parameter constrained
  quadratic and doubly-constrained cubic fits are the most appealing
  models.
\end{enumerate}

\begin{figure}[t]
\includegraphics[width=\linewidth]{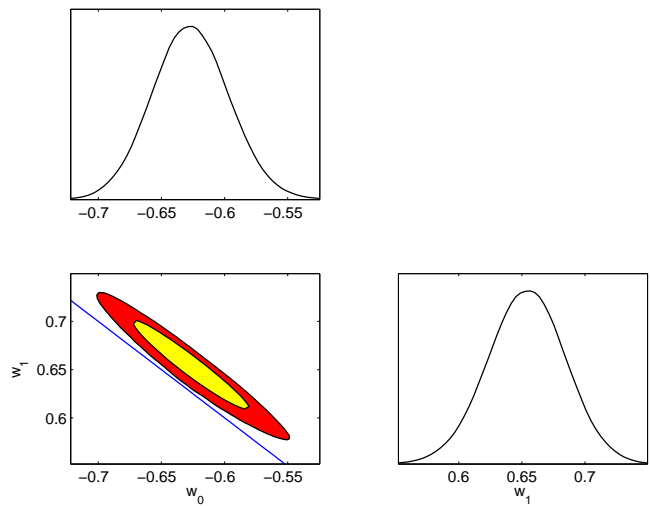}
\caption{Parameter constraints on the linear (CPL)
  parameterization. The constrained models lie on the line $w_1 = -w_0$, 
  shown as the (blue) line on the $w_0$, $w_1$ plot.} 
\label{fig:cplcons}
\end{figure}

\begin{figure}[t]
\includegraphics[width=\linewidth]{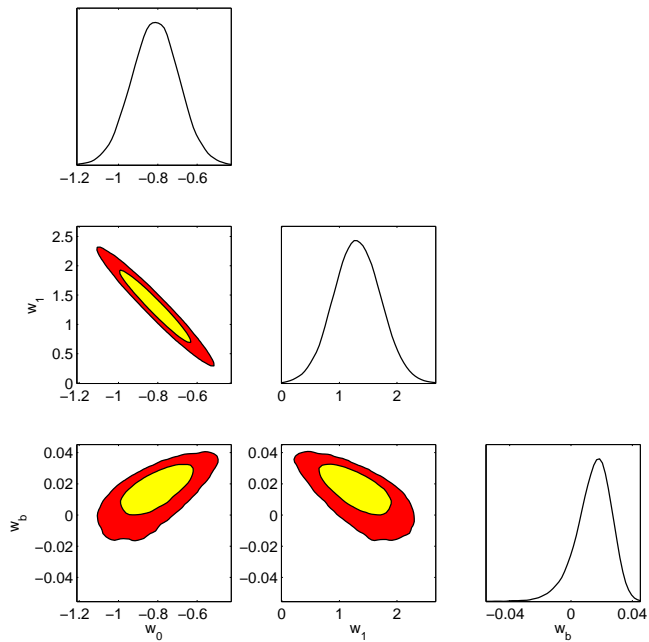}
\caption{Parameter constraints on the quadratic parameterization. Here
  $w_b = w_0+w_1+w_2 $, which is the combination which is held at zero
  in the constrained quadratic case.}
\label{fig:quadcons}
\end{figure}

\begin{figure}[t]
\includegraphics[width=\linewidth]{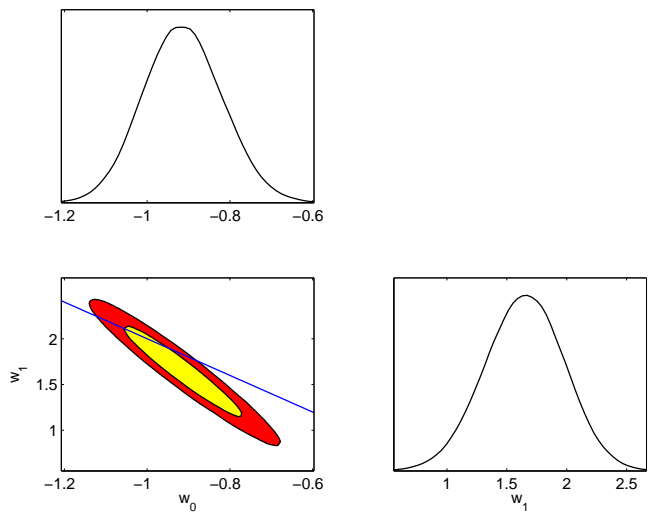}
\caption{Parameter constraints on the constrained quadratic
  parameterization, where $w(a=0)$ is set to zero. The
  doubly-constrained models lie on the line $w_1=-2w_0$, shown as the
  (blue) line on the $w_0$, $w_1$ plot.}
\label{fig:redquadcons}
\end{figure}

More relevant than the best-fit parameters are the ranges of parameter
values permitted by the data. As all constant $w$ models are ruled
out, we show in Figs.~\ref{fig:cplcons} and \ref{fig:quadcons} the
constraints on the parameters of the CPL and quadratic
parameterizations, and in Fig.~\ref{fig:redquadcons} we show the
constraints on the constrained quadratic model, which forces
$w(a=0)=0$.

\begin{figure}[t]
\includegraphics[width=0.90\linewidth]{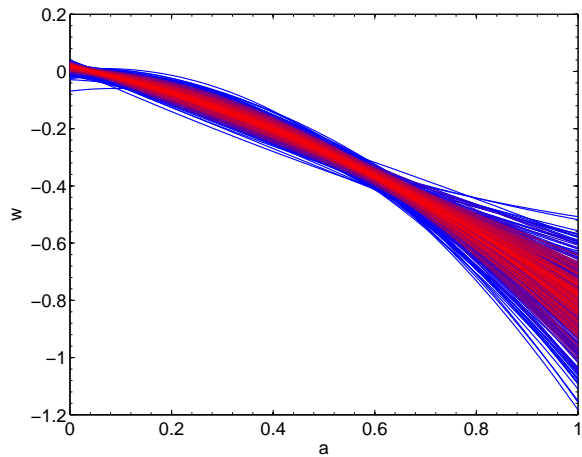}
\caption{An illustration of the form of the best fitting $w(a)$ curves
  for the quadratic case. A sparse sampling of 400 chain elements,
  colour coded by likelihood with the red (lighter) shading the
  highest, is shown.}
\label{fig:wconf}
\end{figure}

For illustrative purposes, Fig.~\ref{fig:wconf} shows 400 quadratic
$w(a)$ curves drawn randomly from the Markov chain.  They are shaded
so that the likelihood increases from blue (darker) to red
(lighter). We can see that the high-redshift region near $a=0$ is
strongly constrained and requires $w\approx0$, while the value of the
total equation of state parameter today, $w_0=w(a=1)$, is only very
weakly constrained by the data sets used in this work.\footnote{As we
  were completing this work, Mortonson et al.~\cite{MHH} arXived a
  paper considering this specific point in much more detail. This
  point had also previously been noted in Ref.~\cite{cor}.} Due to the
integrated nature of the distance constraints on $w$, curves that
oscillate around the best-fit incur only a small penalty. This becomes
more problematic when going to higher order in power law expansions
since then more oscillations become possible. As the oscillations have
to average out, the expansion parameters are highly correlated and not
really independent, which can already be seen for $w_0$ and $w_1$ in
the quadratic case in Fig.~\ref{fig:quadcons}.

We also carried out a different analysis where the equation of state
$w$ is allowed to take different values in binned regions of scale
factor $a$, for simplicity taken to be linearly spaced with 10 or 20
bins. A Principal Component Analysis showed that three modes were well
measured ($\sigma<0.1$), supporting the conclusions above. This
analysis also showed that the best measured mode was peaked in the
highest-redshift bin.

\section{Conclusions}

Due to the dark degeneracy, there is no unique split into dark matter
and dark energy. For this reason, we considered in this paper the
total dark sector equation of state. We parameterized it with
polynomial expansions in the scale factor $a$ and used type Ia
supernovae, baryon acoustic oscillations, and the CMB peak position to
find constraints on the expansion parameters.

The strongest constraints come at very high redshift, from the CMB and
BAO measurements of the sound horizon at decoupling.  There the dark
fluid must evolve with an equation of state close to zero, to recover
the correct angular scale of the acoustic oscillations. At low
redshift the constraints from SN-Ia and BAO are weaker, but require a
negative pressure fluid with $w_{{\rm dark}} \simeq - 0.8$.

We found that $\Lambda$CDM gave the best one-parameter fit to current
data, bettered only by other models with at least two parameters such
as the doubly-constrained cubic expansion. However, none did very much
better than $\Lambda$CDM. Accordingly, $\Lambda$CDM remains the most
compelling interpretation of present data, despite the dark
degeneracy. Nevertheless our polynomial fits indicate the constraints
that can be applied to more general types of dark sector model.

One could speculate about how future measurements of the same general
type might tighten our constraints, but as remarked in the
introduction they cannot address the issue of separability of the dark
sector into components. Rather, to defeat the dark degeneracy what is
needed are \emph{non-gravitational} probes of the sector, for instance
direct detection of dark matter particles at accelerators or in
underground experiments \cite{mydeg}. Such detections would
immediately move such particles from the dark sector to the `known'
sector, removing them from the dark sector degeneracy.

%======================================%
%<<<<<<<<<<< ACKNOWLEDGMENTS >>>>>>>>>>%
%======================================%

\begin{acknowledgments}
M.K., A.R.L., and D.P.\ were supported by STFC (UK).  C.G.\ was
supported by the National Science Foundation of China under the
Distinguished Young Scholar Grant 10525314, the Key Project Grant
10533010, and Grant 10575004; by the Chinese Academy of Sciences under
grant KJCX3-SYW-N2; and by the Ministry of Science and Technology
under the National Basic Sciences Program (973) under grant
2007CB815401.  A.R.L. thanks the Institute for Astronomy, University
of Hawaii, for hospitality while this work was completed. We
acknowledge use of the CosmoMC package \cite{cosmomc}.
\end{acknowledgments}

%======================================%
%<<<<<<<<<<<< BIBLIOGRAPHY >>>>>>>>>>>>%
%======================================%

%%%%%%%%%%%%%%%%%%%%%%%%%%%%%%%%%%%%%%%%%%%%%%%%%%%%%%%%%%%%%%%%%%%%%%%%
\end{document}